\documentclass{elsart}
\usepackage{graphicx}
\usepackage{amsmath}
\usepackage{amssymb}
\usepackage{epsfig}
\usepackage{dcolumn}% Align table columns on decimal point
\usepackage{bm}% bold math
\begin{document}
\begin{frontmatter}

\title{Effective temperature of self--similar time series:
analytical and numerical developments}
\author{Alexander Olemskoi},
\ead{alex@ufn.ru}
\author{Sergei Kokhan}
\address{Department of Physical Electronics,
Sumy State University,\\ Rimskii-Korsakov St. 2, 40007 Sumy, Ukraine}
\date{}

\begin{abstract}
Within both slightly non--extensive statistics and related numerical model, a
picture is elaborated to treat self--similar time series as a thermodynamic
system. Thermodynamic--type characteristics relevant to temperature, pressure,
entropy, internal and free energies are introduced and tested. Predictability
conditions of time series analysis are discussed on the basis of Van der Waals
model. Maximal magnitude for time interval and minimal resolution scale of the
value under consideration are found and analyzed in details. The statistics
developed is shown to be governed by effective temperature being exponential
measure of the fractal dimension of the time series. Testing of the analytical
consideration is based on numerical scheme of non--extensive random walk. A
statistical scheme is introduces to present numerical model as a grand
canonical ensemble for which entropy and internal energy are calculated as
functions of particle number. Effective temperature is found numerically to
show that its value is reduced to averaged energy per one degree of freedom.
\end{abstract}


\end{frontmatter}
%%%%%%%%%%%%%%%%%%%%%%%%%%%%%%%%%%%%%%%%%%%%%%%%%%%%%%%%%%%%%%%%%%%%%%%
\section{Introduction}

Time series analysis allows one to elaborate and verify macroscopic models of
complex systems evolution on the basis of data analysis \cite{0}. This analysis
is known to be focused on numerical calculations of correlation sum for delay
vectors that allow to find principle characteristics of the time series. Being
traditionally a branch of the theory of statistics, time series analysis is
based on the class of models of harmonic oscillator which are related to the
simplest case of the Gaussian random process \cite{A}. But well known a real
time series is relevant rather to the L\'evy stable processes, than the
Gaussian ones being very special case \cite{B}. Because the former are
invariant with respect to dilatation transformation \cite{S}, the problem is
reduced to consideration of self--similar stochastic processes.

The simplest characteristic of a time series is known to be the Lyapunov
exponent the largest of whose positive values yields predictability domain in
the system behavior. A range of complexity in such behavior is determined by
the Kolmogorov--Sinai entropy that equals to sum over positive magnitudes of
whole set of the Lyapunov exponents\footnote{We have in mind hyperbolic
attractors.} and can be reduced to the usual Shannon value in information
theory. With passage to nonlinear system the probability $p_n$ of $n$--th
scenario of the system behavior is transformed into power function $p_n^q$
determined by index $q\leq 1$, so that the Kolmogorov--Sinai entropy should be
replaced by the Renyi one
\begin{eqnarray}
K_q\equiv\frac{1}{1-q}\ln\sum_n p_n^q. \label{000}
\end{eqnarray}
Respectively, governing equation ${\rm d}p/{\rm d}\epsilon=-\beta p^q$,
$\beta={\rm const}>0$ describes probability variation with energy
$\epsilon=\epsilon_n$ to derive the Tsallis distribution \cite{2}
\begin{eqnarray}
p(\epsilon)\propto\left[1-(1-q)\beta\epsilon\right]^{\frac{1}{1-q}}. \label{01}
\end{eqnarray}
In the limit $q\to 1$, this distribution takes usual Boltzmannian form
$p(\epsilon)\propto\exp(-\beta\epsilon)$ falling down exponentially fast in
contrary to the power asymptotic of the generalized exponent (\ref{01}).
Physically, such a behavior is caused by self--similarity of
non--extensive system \cite{3} --- \cite{6}.

As a result, the problem appears to study self--similar time series that
present processes corresponded to power--law distribution type of
Zipf--Mandelbrot ones \cite{M}. This work is devoted to both analytical and
numerical considerations of such type time series as a thermodynamic system. It
is worthwhile to stress that our approach is principally different off the
pseudo--thermodynamic formalism developed for consideration of
multifractal--type objects (see \cite{BS}, \cite{BP} and references therein).
Indeed, if within latter formalism the role of state parameters play related
multifractal indices, we introduce a set of effective state parameters, whose
meaning is of true thermodynamic type (so, an effective temperature is a
measure of data scattering being related to the fractal dimension
of the time series).

The paper consists of two main Sections 2 and 3, the first of which is devoted
to analytical consideration and the second one --- to numerical study. We start
analytical consideration with elaboration of a model which allows us to address
a self--similar time series as slightly non-extensive thermodynamic system (see
Subsection 2.1). Following Subsection 2.2 is based on calculations of both
entropy and internal energy of the time series. As a result,
thermodynamic--type characteristics such as temperature and entropy, volume and
pressure, internal and free energies are introduced. Their testing for the
model of ideal gas is shown to be basis for statistics of self--similar time
series. Subsection 2.3 contains calculations of corrections to the ideal gas
approach when external field and particle interaction are switched on. On the
basis of Van der Waals model, we obtain the expressions for the specific heat
and susceptibility as functions of the temperature. Subsection 2.4 is devoted
to discussion of the physical meaning of the results obtained within the
framework where the predictability of behavior of time series is mimicked as
stability conditions of a non--ideal gas. We find maximal magnitudes for time
interval and minimal resolution scale of the value under consideration.
Subsection 2.5 shows that a temperature governing time series statistics is
exponential measure of a self--similarity index related to fractal dimension of
the time series.

Testing of the analytical consideration is based on numerical scheme of
non--extensive random walk \cite{x(t)} whose stochastic equation and its
solutions are treated in Subsection 3.1. As a result, we obtain non--trivial
time series whose form is governed by a friction coefficient fixing related
fractal dimension. In Subsection 3.2 we introduce a statistical scheme that
allows us to consider modeled time series as a grand canonical ensemble for
which we calculate entropy and internal energy as functions of particle number.
In Subsection 3.3 we find numerically effective temperature and
show that its value is reduced to averaged kinetic energy per one particle of
ideal gas. In concluding Section 4 we confirm numerically exponential
dependence of the effective temperature on the fractal dimension that was found
analytically in Subsection 2.5. We discuss also multifractional and clustering
properties of time series.

Finally, several equalities needed in quoting are placed in Appendices A, B.

\section{Analytical study of self--similar time series}

The principle peculiarity of time series is that it evolves during large enough
but finite time interval ${\mathcal T}<\infty$. On the other hand, the
self--similarity means that a series relates to a fractal manifold
characterized by the fractal dimension $D$. We show below that both properties
pointed out are taken into account in natural way if we shall use non-extensive
statistical mechanics where a simple combination of quantities ${\mathcal T}$
and $D$ fixes the non-extensivity exponent $q$ (see Eqs.(\ref{L1}), (\ref{Q})
below).

\subsection{Description of time series as slightly non-extensive statistical
system}

We start with consideration of $d$--dimensional time series ${\bf x}(t_i)$
related to the set $\{{\bf x}_i\}$ of consequent values ${\bf x}_i\equiv{\bf
x}(t_i)$ of principle variable ${\bf x}(t)$ taken at discrete time instant
$t_i\equiv i\tau$ that we obtain as result of dividing a whole time series
length ${\mathcal T}\equiv N_0\tau$ by $N_0$ equal intervals $\tau$. Following
the ergodic hypothesis we shall imitate the time series ${\bf x}(t_i)$ by
the set of generalized coordinates $\{{\bf x}_n\}$ supplemented
by conjugated set $\{{\bf v}_n\}$ of
velocities that show jumping rates of coordinates ${\bf x}_i$ with
the time variation. It is principally important to take into account that
mapping of the time series ${\bf x}(t_i)$ into the manifold $\{{\bf x}_n\}$ is
not mutually single valued because different terms ${\bf x}(t_i)$ and ${\bf
x}(t_j)$ may be equal (similarly, it appears at mapping of velocity time
series ${\bf v}(t_i)$ into related manifold $\{{\bf v}_n\}$). This
peculiarity is displayed first of all in that the number $N_0$ of specimens of
the time series is much more than the total number $N$ of terms of related
manifold. In the following, just the latter number $N$ plays the role of the particle
number of statistical system. Obviously, this system should be considered as
grand canonical ensemble with variable number $N_n$ of particles in a state
$n$.

In the simplest case of Markovian consequence, one has the velocity ${\bf
v}_n\equiv ({\bf x}_{n}-{\bf x}_{n-1})/\tau$. For more complicated series with
$m$-step memory, the velocity magnitude is defined as follows:
\begin{eqnarray}
v_n^{(m)}\equiv\sqrt{{1\over m}\sum_{l=1}^{m}{\vec \delta}_l^2(m, n)},\qquad
{\vec\delta}_l(m,n)\equiv{{\bf x}_{(n-m)+l}-{\bf x}_{(n-m)+(l-1)}
\over\tau}.
\label{A}
\end{eqnarray}

Along the line of the ergodic hypothesis,
the paradigm of our approach is to address the time series as a physical
system defined by
an effective Hamiltonian ${\mathcal H}={\mathcal H}\{{\bf x}_n, {\bf v}_n\}$
on whose basis statistical characteristics of this series could be found.
If one proposes that series terms ${\bf x}_n$ related to different $n$
are not connected, the effective Hamiltonian is additive:
\begin{eqnarray}
{\mathcal H}=\sum\limits_{n=1}^{N}\varepsilon_n N_n,\quad
\varepsilon_n\equiv\varepsilon({\bf x}_n, {\bf v}_n). \label{B}
\end{eqnarray}
Physically, this means that the series under consideration is relevant to an
ideal gas comprising of $N$ identical particles with energy $\varepsilon_n$ and
number $N_n$ in state $n$. Further, we suppose different terms of time series
to be statistically identical, so that effective particle energy does not
depend on coordinate ${\bf x}_n$: $\varepsilon({\bf x}_n, {\bf v}_n)
\to\varepsilon({\bf v}_n)$. Moreover, since this energy does not vary with
inversion of the coordinate jumps ${\bf x}_{n}-{\bf x}_{n-1}$, the function
$\varepsilon({\bf v}_n)$ should be even. We use the simplest square form
\begin{eqnarray}
\varepsilon_n={1\over 2}{\bf v}_n^2,
\label{C}
\end{eqnarray}
which is reduced to the usual kinetic energy for a particle with
mass $1$.
With switching on an external force ${\bf F}={\bf const}$, particle energy (\ref{C})
becomes as follows:
\begin{eqnarray}
\varepsilon_n={1\over 2}{\bf v}_n^2-{\bf F}{\bf x}_n.
\label{CC}
\end{eqnarray}
Finally, when time series has a microscopic memory,
dimension of the delay vectors ${\vec \delta}_i(m, n)$ in the definition (\ref{A}) needs taking $m>1$.
Moreover, if time series terms ${\bf x}_m$, ${\bf x}_n$ with $m\ne n$ are clustered, the Hamiltonian
becomes relevant to a non--ideal gas with interaction $w_{mn},~m\ne n$:
\begin{eqnarray}
{\mathcal H}={1\over 2}\sum\limits_{n=1}^{N}{\bf v}_n^2+
{1\over 2}\sum\limits_{m\ne n}w_{mn}.
\label{CCC}
\end{eqnarray}

To study a behavior of the time series as a whole one needs to fulfill
summation over a set of states given by manifold $\{{\bf x}_n, {\bf v}_n\}$
that is relevant to the system phase space. In so doing, it is convenient to
pass to related integrations as following:
\begin{eqnarray}
\sum\limits_{\{{\bf x}_n, {\bf v}_n\}}\Rightarrow
\iint\prod\limits_{n=1}^{N}{{\rm
d}{\bf x}_n{\rm d}{\bf v}_n\over N!\Delta}=
{\mathcal  N}^{-1}\prod\limits_{n=1}^{N}
\iint{\rm d}{\bf y}_n{\rm d}{\bf u}_n.
\label{D1}
\end{eqnarray}
Here, the factorial takes into account statistical identity of
the time series terms, $\Delta$ is effective Planck constant that determines a
minimal volume of the phase space
per a particle related to a term. The inverted factor
\begin{eqnarray}
{\mathcal  N}\equiv N!\left({X^2\over\tau\Delta}\right)^{-dN}\simeq
\left[{{\rm e}X^{2d}\over N(\tau\Delta)^d}\right]^{-N}
\label{D2a}
\end{eqnarray}
is caused by change of variables
\begin{eqnarray}
{\bf y}_n\equiv{{\bf x}_n\over X},\quad
{\bf u}_n\equiv{\tau{\bf v}_n\over X}
\label{D2b}
\end{eqnarray}
rescaled with respect to macroscopic length $X$
being chosen to guarantee the conditions
\begin{eqnarray}
\int{\rm d}{\bf y}_n=1,\quad
\int{\rm d}{\bf u}_n=1.
\label{D2c}
\end{eqnarray}

According to Introduction self--similarity condition forces to use a statistics
type of given by the Tsallis--type distribution (\ref{01}). However, the latter
is appeared to be inconvenient because the condition
\begin{eqnarray}
\sum\limits_n p_n^q\equiv\left<1\right>_q\ne 1
\label{Da}
\end{eqnarray}
takes place and definition of the internal energy reads:
\begin{eqnarray}
E=\sum\limits_n\frac{\varepsilon_n p_n^q}{\left<1\right>_q}. \label{Db}
\end{eqnarray}
To take into account these constraints an escort distribution was proposed to
use \cite{7}
\begin{eqnarray}
{\mathcal  P}_n\equiv\frac{p_n^q}{\left<1\right>_q}.
\label{Dc}
\end{eqnarray}
In explicit form it reads as follows:
\begin{eqnarray}
{\mathcal P}_q\{{\bf y}_n,{\bf u}_n\}=\left\{
\begin{array}{ll}
{1\over Z}\left[1-(1-q){{\mathcal  H}\{{\bf y}_n, {\bf u}_n\}-E\over
\left<1\right>_q T_s}\right]^{q\over 1-q}\ {\rm at}\
(1-q){{\mathcal H}\{{\bf y}_n, {\bf u}_n\}-E\over\left<1\right>_q T_s}<1,\\
\\
0\quad\quad\quad\quad\quad\quad\qquad\quad
\quad\quad\quad\quad\quad\quad\qquad\quad{\rm otherwise}.
\end{array} \right.
\label{D}
\end{eqnarray}
Here, the partition function is defined by the condition
\begin{eqnarray}
Z\equiv{\mathcal N}^{-1}\prod\limits_{n=1}^{N}\iint\left[1-(1-q)
{{\mathcal H}\{{\bf y}_n,{\bf u}_n\}-E\over
\left<1\right>_q T_s}\right]^{q\over 1-q}
{\rm d}{\bf y}_n{\rm d}{\bf u}_n,
\label{E}
\end{eqnarray}
where $0<q<1$ is a parameter of non-extensivity, $T_s$ is energy scale.
Internal energy $E$ is determined by the equality
\begin{eqnarray}
E\equiv {\mathcal  N}^{-1}\prod\limits_{n=1}^{N}\iint{\mathcal H}
\{{\bf y}_n, {\bf u}_n\}{\mathcal  P}_q\{{\bf y}_n, {\bf u}_n\}
{\rm d}{\bf y}_n{\rm d}{\bf u}_n,\quad
\label{F}
\end{eqnarray}
and normalization parameter $\left<1\right>_q$ is expressed by
the partition function (\ref{E}) in accordance with Eq.(\ref{G})
in Appendix A.

To check the statistical scheme
proposed let us address firstly trivial case of time series
${\bf x}_n={\rm{\bf const}}$.
Here, the particle energy $\varepsilon$ is a constant as well,
so that the Hamiltonian is ${\mathcal  H}=N\varepsilon$.
The partition function $Z={\mathcal  N}^{-1}$ and the
normalization parameter
$\left<1\right>_q={\mathcal  N}^{-(1-q)}$
are given by inverted normalization factor (\ref{D2a}), while
the internal energy $E=N\varepsilon$ is reduced to the Hamiltonian.
Then, the entropy $H=-a\ln{\mathcal N}$, $a=\frac{1}{2}(1-q)dN\ne 0$
obtained according to definition
(\ref{H1}) given in Appendix A is reduced to zero if only the normalization
factor takes the value ${\mathcal N}=1$.
As a result, we find effective Planck constant:
\begin{eqnarray}
\Delta=\left({{\rm e}\over N}\right)^{1\over d}{X^2\over\tau}.
\label{I}
\end{eqnarray}

Our future consideration is stated on the assumption that the volume
$V\equiv X^d$ of $d$-dimensional domain of the ${\bf x}_n$ coordinate variation
in dependence of the particle number $N$ is governed by L\'evy--type law
\begin{eqnarray}
X^d=x^d N^{1\over z}.
\label{J}
\end{eqnarray}
Here, $x$ is a microscopic scale and a dynamic exponent $z$ is reduced to the
fractal dimensionality $D$ of self--similar manifold \cite{3a}
\begin{eqnarray}
z=D.
\label{Jjj}
\end{eqnarray}
Then, one obtains the following scaling relation for the phase space volume
per a term of the time series:
\begin{eqnarray}
\Delta^d={\rm e}\left({x^2\over\tau}\right)^{d}N^{{2\over D}-1}.
\label{K}
\end{eqnarray}
In the case of Gaussian scattering, when $D=2$, the minimal volume
$\Delta^d$ of the phase space does not depend on number $N$
of the time series terms.
Such a condition approves our choice of the relation (\ref{J})
for the whole volume $V\equiv X^d$ as function of the number $N$
of time series terms.

\subsection{Non-extensive thermodynamics of time series as an ideal gas}

Calculations of main thermodynamic quantities of non-extensive ideal gas
arrives at the following expressions for the partition function (\ref{E}) and
the internal energy (\ref{F}) (see \cite{7} --- \cite{7a})
\begin{eqnarray}
Z={V^N\gamma(q)\over N!}\left[{\theta\left<1\right>_q\over 1-q}\right]^{dN\over
2}\left[1+(1-q){dN\over 2}\right]^{-1}\nonumber\\
\cdot\left[1+(1-q){~E\over\left<1\right>_q T_s}\right]^{{1\over 1-q}+{dN\over 2}},
\label{S5}
\end{eqnarray}
\begin{eqnarray}
E={dN\over 2}{V^N\gamma(q)T_s\over N!}\left[{\theta\left<1\right>_q\over
1-q}\right]^{dN\over 2}
\left[1+(1-q){dN\over 2}\right]^{-1}\nonumber\\
\cdot\left[1+(1-q){~E\over\left<1\right>_q T_s}\right]^{{1\over 1-q}+{dN\over
2}}Z^{-q}.\quad
\label{S6}
\end{eqnarray}
Here, $d$-dimensional gas in volume $V\equiv X^d$ is
addressed and the notations are introduced
\begin{eqnarray}
\theta\equiv{2\pi T_s\over\Delta^2},\quad
\gamma(q)\equiv{\Gamma\left({1\over 1-q}\right)
\over\Gamma\left({1\over 1-q}+{dN\over 2}\right)}
\label{M}
\end{eqnarray}
to be determined by the Euler $\Gamma$--function.
Combination of equalities (\ref{S5}), (\ref{S6}) with relation (\ref{G})
yields explicit expression for the normalization parameter:
\begin{eqnarray}
\left<1\right>_q=\left\{ {X^{dN}\gamma(q)\over N!}
\left[{\theta(1+ a)^{{q\over a}+1}\over 1-q}\right]^{dN\over 2}
\right\}^{1-q\over 1- a},
\label{L}
\end{eqnarray}
\begin{eqnarray}
a\equiv\frac{1}{2}(1-q)dN.
\label{L1}
\end{eqnarray}
In the limits
\begin{eqnarray}
1-q\ll 2/d,\quad
N\gg 1
\label{N1}
\end{eqnarray}
when
\begin{eqnarray}
\gamma(q)\simeq\left[{\rm e}(1-q)\right]^{dN\over 2}(1+ a)^{-{1+a\over 1-q}},
\label{N}
\end{eqnarray}
one obtains for the first of entropies (\ref{H1}):
\begin{eqnarray}
H\simeq{Na\over 2(1- a)}\ln\left[{\rm e}^{2+d}\theta^d\left({X^d\over N}\right)^{2}\right].
\label{O}
\end{eqnarray}
With accounting scaling relation (\ref{J}), this expression takes
the usual form
\begin{eqnarray}
H=N{D-1\over D}\ln\left({G\over N}\right),\quad
G\equiv(2\pi{\rm e}T_s)^{dD\over 2}\left(x\over\tau\right)^{-dD}
\label{P}
\end{eqnarray}
if the dynamic exponent is determined as
\begin{eqnarray}
z\equiv D={1\over 1-a}.
\label{Q}
\end{eqnarray}
Respectively, the internal energy (\ref{S6}) and the normalization
parameter (\ref{L}) read:
\begin{eqnarray}
E={dN\over 2}\left({G\over N}\right)^{2a\over d}T_s,\quad
\left<1\right>_q=\left({G\over N}\right)^{2a\over d};\quad
a\equiv{D-1\over D}.\quad
\label{S}
\end{eqnarray}

The physical temperature is defined as follows \cite{9}
\begin{eqnarray}
T\equiv\left<1\right>_q T_s=\left({G\over N}\right)^{2a\over d}T_s
\label{Ss}
\end{eqnarray}
where the last equality takes into account the second of relations (\ref{S}).
This definition guarantees the equipartition law
\begin{eqnarray}
E=CT,\quad C\equiv cN,\quad c\equiv{d\over 2}
\label{1}
\end{eqnarray}
where the quantity
\begin{eqnarray}
C={\partial E\over\partial T}
\label{1a}
\end{eqnarray}
is the specific heat.
It is easily to convince that equations (\ref{P}) --- (\ref{Ss})
arrive at standard thermodynamic relation
\begin{eqnarray}
{\partial H\over\partial E}\equiv{1\over T}.
\label{33}
\end{eqnarray}

Above used treatment is addressed to
a fixed value of the internal energy $E$ \cite{7a}.
In alternative case when the principle state parameter is the temperature $T$,
we should pass to the conjugate formalism \cite{7}. Here, standard definition
\begin{eqnarray}
F\equiv E-TH
\label{3aa}
\end{eqnarray}
of the free energy arrives at the dependence
\begin{eqnarray}
F=-CT\ln\left({~T\over{\rm e}T_s}\right).
\label{Zz}
\end{eqnarray}
Then, the thermodynamic identity
\begin{eqnarray}
{\partial F\over\partial T}\equiv - H
\label{3}
\end{eqnarray}
yields the relation
\begin{eqnarray}
H=C\ln\left({T\over T_s}\right)
\label{Zz1}
\end{eqnarray}
that plays a role of the heat equation of states.
It arrives at the usual definition of the specific heat (cf. Eq. (\ref{1a}))
\begin{eqnarray}
C=T{\partial H\over\partial T}.
\label{1aa}
\end{eqnarray}

Let us introduce now a specific entropy per unit time
\begin{eqnarray}
h\equiv(d\tau)^{-1}{{\partial}H\over{\partial}N}=
\tau^{-1}H_1-r
\label{S1}
\end{eqnarray}
to be determined by a minimal entropy
\begin{eqnarray}
H_1=(D-1)\left[\ln\sqrt{2\pi{\rm e}T_s}-\ln\left(x\over\tau\right)\right]
\label{S2}
\end{eqnarray}
and a redundancy
\begin{eqnarray}
r=\frac{a}{d\tau}\ln({\rm e}N),\quad
a\equiv\frac{D-1}{D}.
\label{S3}
\end{eqnarray}
Dependencies on the scale $x$
\begin{eqnarray}
h(x)={\rm const}-{D-1\over\tau}\ln\left({x\over\tau}\right),\quad
r(x)={\rm const}
\label{S4}
\end{eqnarray}
notice that the system behaves in a stochastic manner \cite{10}.

Effective pressure is defined as
\begin{eqnarray}
p\equiv-\tau{\partial h\over\partial x}
\label{3a}
\end{eqnarray}
to measure specific entropy variation with respect to the time series scale.
Then, we arrive at a mechanic--type equation of states
\begin{eqnarray}
px=D-1
\label{3b}
\end{eqnarray}
being additional to the relation (\ref{Zz1}) of entropy to temperature.
According to Eq. (\ref{3b}), definition of the pressure coefficient
\begin{eqnarray}
\kappa\equiv{\partial p\over\partial\left(x^{-1}\right)}
\label{3c}
\end{eqnarray}
shows that it is fixed by the dynamic exponent (\ref{Q}) being the fractal
dimensionality of self--similar time series:
\begin{eqnarray}
\kappa=D-1.
\label{3cc}
\end{eqnarray}
It is worthwhile to note that effective pressure (\ref{3a}) is
introduced as derivative of the specific entropy $h$ with respect
to the microscopic scale $x$.
That reduces a susceptibility of the time series
to the pressure coefficient (\ref{3c}) inverted.

\subsection{Corrections to the ideal gas approach}

Let us focus now on effect of both external field and particle interaction
whose switching on is expressed by equalities (\ref{CC}), (\ref{CCC}). An
external force ${\bf F}={\bf const}$ causes the second term in Hamiltonian
(\ref{CC}) to arrive at the factor
\begin{eqnarray}
Z_{ext}=Z_d\left[{\sinh\left({FX\over 2T}\right)\over
{FX\over 2T}}\right]^{N}
\label{G1}
\end{eqnarray}
in partition function (\ref{S5}). Here, $Z_d$ is a factor depended on the
dimension $d$ only and we put $q\to 1$ due to the conditions (\ref{N1}).
According to the definition (\ref{H1}) the factor (\ref{G1}) yields the entropy
addition
\begin{eqnarray}
H_{ext}= Na\ln\left[{\sinh\left({FX\over 2T}\right)
\over{FX\over 2T}}\right]
\label{G2}
\end{eqnarray}
where we suppress unessential term. With increasing homogeneous external field,
the entropy $H_{ext}$ grows quadratically at $FX\ll T$ and linearly at $FX\gg
T$.

According to \cite{11}, cluster expansion of
the particle interaction in Hamiltonian (\ref{CCC})
results in additional factor in partition function (\ref{S5}):
\begin{eqnarray}
&Z_{int}= 1 - {N^2\over 2V}\left(v+{w\over T}\right);&
\label{G3}\\
&w\equiv S_d\int\limits_\epsilon^\infty w(x)x^{d-1}{\rm d}x,~
S_d\equiv{2\pi^{d/2}\over\Gamma(d/2)};~~
v\equiv{S_d\over d}\epsilon^d&
\nonumber
\end{eqnarray}
where $\epsilon$ is effective radius of the particle core.
Relevant entropy addition
\begin{eqnarray}
H_{int}=-{aN^2\over 2V}\left(v+{w\over T}\right)
\label{G4}
\end{eqnarray}
decreases monotonically with growth of the particle volume $v$. Increase of the
temperature $T$ causes the entropy decrease for attractive interaction $w<0$
and its increase in the case of repelling one $(w>0)$.

Entropy additions (\ref{G2}), (\ref{G4}) arrive at
total value of the specific heat (\ref{1aa}) in the following form:
\begin{eqnarray}
C={d\over 2}N-aN\left\{\left[{FX\over 2T}{\rm coth}\left({FX\over 2T}\right)
-1\right]- {N\over 2V}{w\over T}\right\}\quad
\label{Gg}
\end{eqnarray}
where formula (\ref{1}) are taken into account. On the other hand, making use
of equalities (\ref{S1}), (\ref{3a}), (\ref{G2}), (\ref{G4}) arrives at the
pressure addition $\delta p$ determined by the equality
\begin{eqnarray}
\delta p\cdot x=-\frac{a}{d}\left[{FX\over 2T}{\rm coth}\left({FX\over 2T}\right)
-1\right]+a\frac{N}{V}\left(v+{w\over T}\right).\quad
\label{333a}
\end{eqnarray}
Then, with accounting (\ref{3cc}) the pressure coefficient (\ref{3c})
takes the form
\begin{eqnarray}
\kappa={a\over 1-a}
+{a\over d}\left\{1+\left[{{FX\over 2T}\over{\cosh\left({FX\over 2T}\right)}}
\right]^2\right\}+(1+d)an\left(v+{w\over T}\right)
\label{333}
\end{eqnarray}
where effective density
\begin{eqnarray}
n\equiv{N\over V}=x^{-d}N^a
\label{a333}
\end{eqnarray}
is introduced.
In the limiting case of low temperatures $T\ll FX/2$, we obtain:
\begin{eqnarray}
C\simeq{d\over 2}N-{aN\over 2T}\left(FX-nw\right),
\label{Ggg}
\end{eqnarray}
\begin{eqnarray}
\kappa\simeq{a\over 1-a}+{a\over d}\left[1+\left({FX\over T}\right)^2
\exp\left(-{FX\over T}\right)\right]+(1+d)an\left(v+{w\over T}\right).
\label{333b}
\end{eqnarray}
Respectively, in opposite case $T\gg FX/2$, one has:
\begin{eqnarray}
C\simeq{d\over 2}N+{aN\over 2T}\left[nw-
{\left(FX\right)^2\over 6T}\right],
\label{Gggg}
\end{eqnarray}
\begin{eqnarray}
\kappa\simeq{a\over 1-a}
+{a\over d}\left[1+{1\over 4}\left({FX\over T}\right)^2\right]+
(1+d)an\left(v+{w\over T}\right).
\label{333bb}
\end{eqnarray}
Thus, in the limits of both low and high temperatures, the influence of
external field $F$ reduces to hyperbolically decreasing addition $(FX/T)^n$,
$n=1, 2$ to the specific heat $C$; accordingly, the pressure coefficient
$\kappa$ gets the constant $a/d$ only. In a similar manner, the particle
interaction affects hyperbolically on the specific heat, while the pressure
coefficient takes the term being proportional to the particle volume $v$.

\subsection{Stability conditions}

It might seem the analysis of a time series would appeared to be the subject of
more much study if both the interval $\tau$ and the scale $x$ take infinitely
decreasing magnitudes (respectively, the number $N$ tends to infinitely large
values). However, we will show now that external influence and term clustering
arrive at predictability boundaries in behavior of relevant system that are
represented as instability of modeling thermodynamic system. Our analysis is
stated on stability conditions
\begin{eqnarray}
C>0,\quad \kappa>0
\label{3d}
\end{eqnarray}
for the specific heat $C$ and the pressure coefficient $\kappa$ given by
equalities (\ref{Gg}) and (\ref{333}).

In the simplest approach being modeled by the ideal gas in Subsection 2.2, we
arrives at the natural restriction for the dynamic exponent:
\begin{eqnarray}
z\equiv D>1.
\label{3e}
\end{eqnarray}
It means non-extensive dependence of the system volume (\ref{J}) on the
particle number $N$. On the other hand, in accordance with relation (\ref{Q})
the magnitude of the principle index (\ref{L1}) is limited by condition
\begin{eqnarray}
a<1
\label{3ee}
\end{eqnarray}
which restricts the rate of increasing effective density (\ref{a333})
with $N$.

As is seen from expressions (\ref{Gg}), (\ref{Ggg}), (\ref{Gggg}), the specific
heat falls down monotonically with temperature decrease. To analyze such a
behavior quantitatively let us consider the case of low temperatures $T\ll
FX/2$, when we can use the simplest dependence (\ref{Ggg}). Then, it is easily
to see that at temperature less than critical magnitude
\begin{eqnarray}
T_c={a\over d}\left(FxN^{1-a\over d}-wx^{-d}N^a\right)
\label{Gg1}
\end{eqnarray}
the system becomes nonstable due to external force $F$ and
interparticle attraction $w<0$.
If the number $N$ is much less than critical value $N_c$ defined as
\begin{eqnarray}
N_c^{{1\over d}-{1+d\over d}a}={|w|\over F}x^{-(1+d)}, \label{Gg2}
\end{eqnarray}
main contribution gets interparticle attraction $w<0$,
whereas in opposite case $N\gg N_c$ --- external force $F$.
With passage to more complicated case
of high temperatures $T\gg FX/2$, when one follows to use
temperature dependence (\ref{Gggg}), the physical situation does not
change qualitatively.

As show estimations (\ref{333b}), (\ref{333bb}),
the pressure coefficient becomes negative if microscopic scale $x^d$
is less than a critical magnitude
\begin{eqnarray}
x_c^d={d(1+d)\over 1+{d\over 1-a}}\left({|w|\over T}-v\right)N^{a}
\label{343}
\end{eqnarray}
that is determined at attractive nature of interaction only $(w<0)$. If both
parameters $w$ and $v$ takes non--zeroth magnitudes, the temperature values are
characterized by another minimal magnitude $T_v=|w|/v$, lower which the
microscopic scale $x_c$ becomes principle. This temperature is less than
critical value (\ref{Gg1}) if elementary volume takes magnitudes lower than
value
\begin{eqnarray}
v_c\equiv{d\over a}n^{-1}={d\over a}x^d N^{-a}
\label{343a}
\end{eqnarray}
where we take into account definition (\ref{a333}).

In the limits $v\to 0$, $F\to 0$, boundary magnitudes (\ref{Gg1}),
(\ref{343}) arrive at appearance of the critical value
\begin{eqnarray}
a_c={(1+d)(1+d^2)\over 2}\left[1-\sqrt{1-{4d^2\over (1+d)(1+d^2)^2}}\right].\ \
\label{343b}
\end{eqnarray}
Relevant fractal dimension of the time series (\ref{Q}) grows monotonically
with dimension of the time series $d$ from the magnitude $D_c=1+\sqrt{2}\simeq
2.414$ at $d=1$ to $D_c\simeq d^2$ at $d\gg 1$ (see Figure 1). As a result, the
critical value (\ref{343b}) arrives at the upper boundary of the fractal
dimensionality $D_c$ being more than topologic magnitude $D_t=2d$ always
($D_c>D_t$). Thus, we can conclude the upper boundary $D_c$ of application of
our approach is appeared to be inessential at all.
\begin{figure}[htb]
\centering
\includegraphics[width=65mm]
{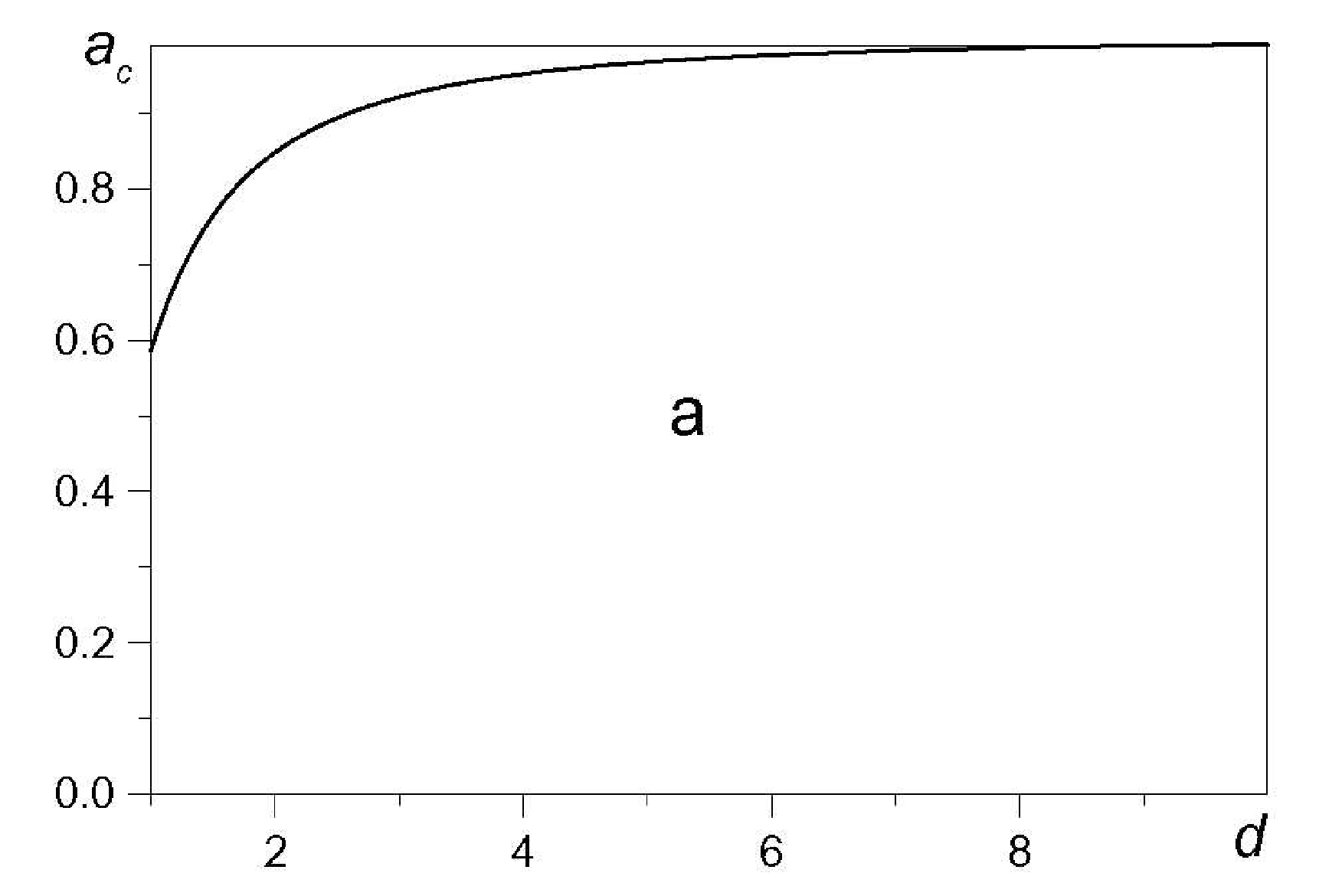}
\includegraphics[width=65mm]
{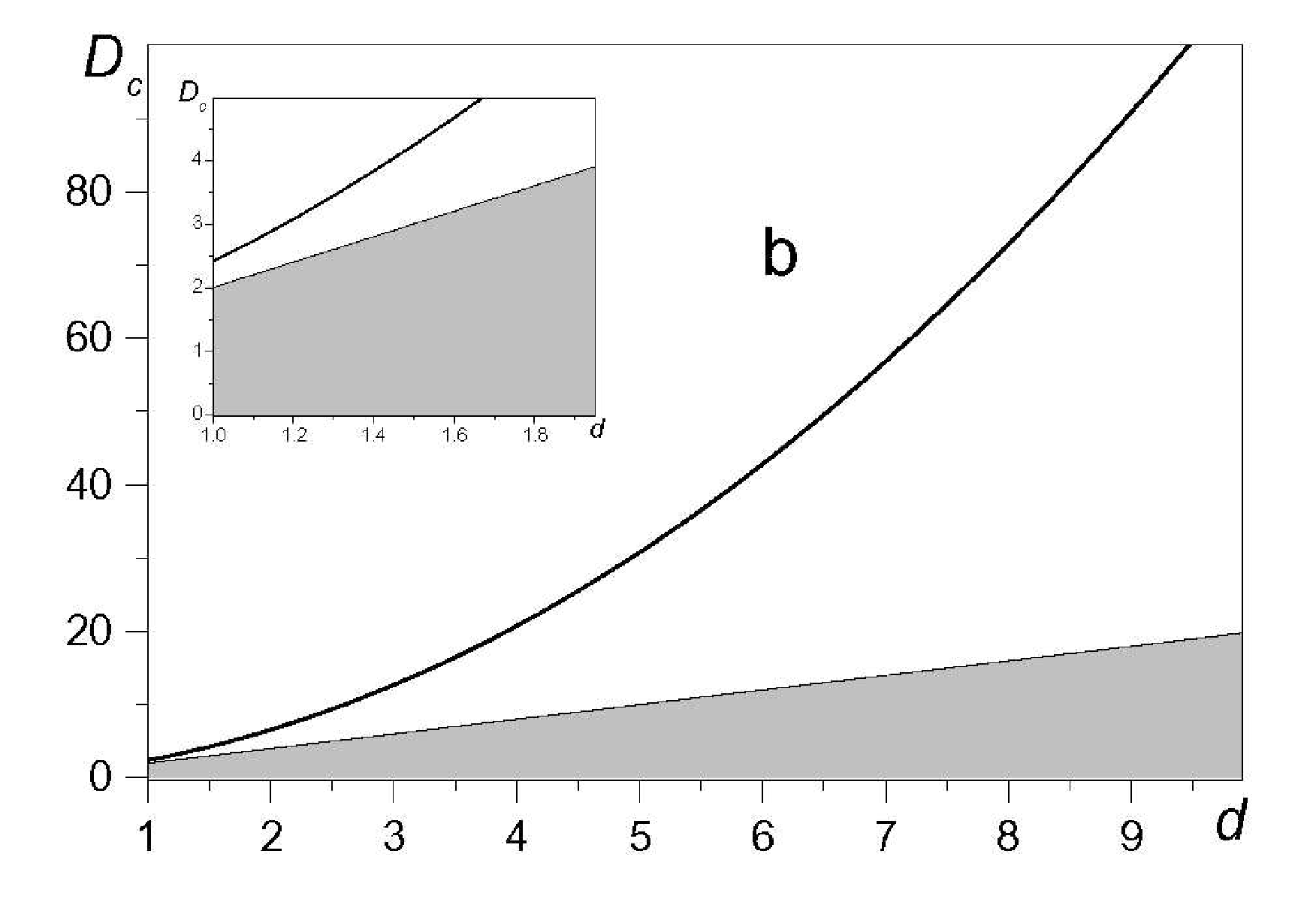} \caption{a. The dependence of the critical value (\ref{343b}) on the
time series dimensionality. b. The same for related maximal value (\ref{Q}) of
the fractal dimension of the time series (solid line) and topologic magnitude
$D_t=2d$ (lower line). The physical domain is shaded.}
\end{figure}

\subsection{Effective temperature of time series}

We have considered above the simplest model which have allowed us to examine
analytically a self--similar time series in standard statistical manner.
Characteristic peculiarity of related equalities is a scale invariance with
respect to variation of the non-extensivity parameter $1-q$ which is contained
everywhere through the parameter (\ref{L1}) that is related to the dynamic
exponent $z$ and the fractal dimension $D$ of the time series according to Eq.
(\ref{Q}). This invariance is clear to be caused by self--similarity of the
system under consideration. For a given time series, the value $D$ is fixed if
it is addressed to a monofractal manifold and takes a closed set of magnitudes
in the case of self--similar system relevant to a multifractal.

In real time series the property of self--similarity can be broken, so that a
dependence on the non-extensivity parameter itself could appear to be very
weak. However, above introduced set of pseudo--thermodynamic characteristics of
time series is kept as applicable and accustomed thermodynamic relations
(\ref{1}) --- (\ref{3aa}), (\ref{3}), (\ref{1aa}), (\ref{3a}) and (\ref{3c})
can be applied to analysis of arbitrary time series.

Main progress in our consideration is that time series statistics
is governed completely by the temperature (\ref{Ss}).
With accounting the second of equalities (\ref{P})
and rescaling the temperature unit $T_s$
into $T_{sc}\equiv (2\pi{\rm e})^a T_s$
we derive to the expression
\begin{eqnarray}
{T\over T_{sc}}=\left[\left({\tau\over X}\right)^2
T_{sc}\right]^{a\over 1-a}.
\label{26}
\end{eqnarray}
Being independent of the number of terms $N$, the time series
temperature shows exponential dependence on the index (\ref{L1})
located under critical magnitude (\ref{343b}).
To establish a character of the power dependence on the ratio of the range $X$ of the
principle variable to the time interval $\tau$, it is naturally to
choice measure units of the temperature in the following manner:
\begin{eqnarray}
T_{sc}\equiv{\rm e}\left({X\over\tau}\right)^2,\qquad T_s\equiv{{\rm
e}^{1-a}\over(2\pi)^a}\left({X\over\tau}\right)^2. \label{27}
\end{eqnarray}
Then, the expression (\ref{26})
for the time series temperature takes the simplest form
\begin{eqnarray}
T=\left({X\over\tau}\right)^2{\rm e}^D,\quad
D\equiv\frac{1}{1-a}
\label{28}
\end{eqnarray}
according to which the value $T$ is the exponential measure of the fractal
dimensionality $D$ of the self--similar time series.

According to consideration given in Subsection 2.4, time series subjected to
external influence and term clustering is limited in predictability that is
emerged as instability of modeling thermodynamic system. On the basis of Van
der Waals model, we have found minimal magnitudes (\ref{343}), (\ref{343a}) for
resolution scale $x_c$ of the value under consideration. Respectively, making
use of the critical temperature (\ref{Gg1}) and the definition (\ref{28}) shows
that a time series is predictable if the microscopic time interval $\tau$ takes
magnitudes less than a critical magnitude $\tau_c$. In the case $N\ll N_c$,
where the boundary magnitude is determined by Eq. (\ref{Gg2}), we find
\begin{eqnarray}
\tau_c=\left({a\over d}|w|\right)^{-{1\over 2}}{\rm e}^{D\over 2}
x^{1+{d\over 2}}
N^{{2+d\over 2Dd}-{1\over 2}}.
\label{Gg31}
\end{eqnarray}
In opposite case $N\gg N_c$, one obtains
\begin{eqnarray}
\tau_c=\left({a\over d}F\right)^{-{1\over 2}}{\rm e}^{D\over 2}
x^{1\over 2}N^{1\over 2Dd}.
\label{Gg3}
\end{eqnarray}
It is easily to convince that the entropy (\ref{Zz1}) takes positive values
within the stability domain $\tau<\tau_c$.

\section{Numerical study of time series}

In this Section, we are aimed to verify numerically the definitions based on
expressions (\ref{A}) --- (\ref{C}), (\ref{D}) --- (\ref{F}), (\ref{L1}),
(\ref{P}) --- (\ref{1aa}). As above analysis has shown, one of peculiarities of
the statistical system under consideration is that it is slightly
non--extensive due to condition (\ref{N1}). Thus, we can put $q=1$ in the
following numerical consideration.

\subsection{Simulations of time series}

In our consideration, we shall follow the numerical scheme of non-extensive
random walk \cite{x(t)} based on the discrete stochastic equation
\begin{equation}\label{discret}
x_{i+1}=\sqrt{\tau}\zeta_i +\left[(1-\gamma\tau)+\sqrt{\tau}\xi_i\right]x_i.
\end{equation}
Here, discrete time $t_i=i\tau$ is fixed by integers $i=0, 1, ...,N_0$ and
minimal interval $\tau$; $\zeta_i$ and $\xi_i$ are additive and multiplicative
stochastic sources normed with white--noise conditions
$\langle\zeta_i\zeta_{j}\rangle=\langle\xi_i\xi_{j}\rangle=\delta_{ij}$;
friction coefficient $\gamma$ determines parameter $\nu=\gamma/(1+\gamma)$
which fixes, in accordance with stationary distribution, non--extensivity
parameter $q\equiv (2-\nu)^{-1}=(1+\gamma)/(2+\gamma)$. Making use of iteration
procedure (\ref{discret}) arrives at stochastic time series, whose form is
shown in Figure 2 for total ${\mathcal T}=500$ and minimal $\tau=0.01$ time
intervals and different $\nu$ values.
\begin{figure}[htb]
\centering
\includegraphics[width=130mm]
{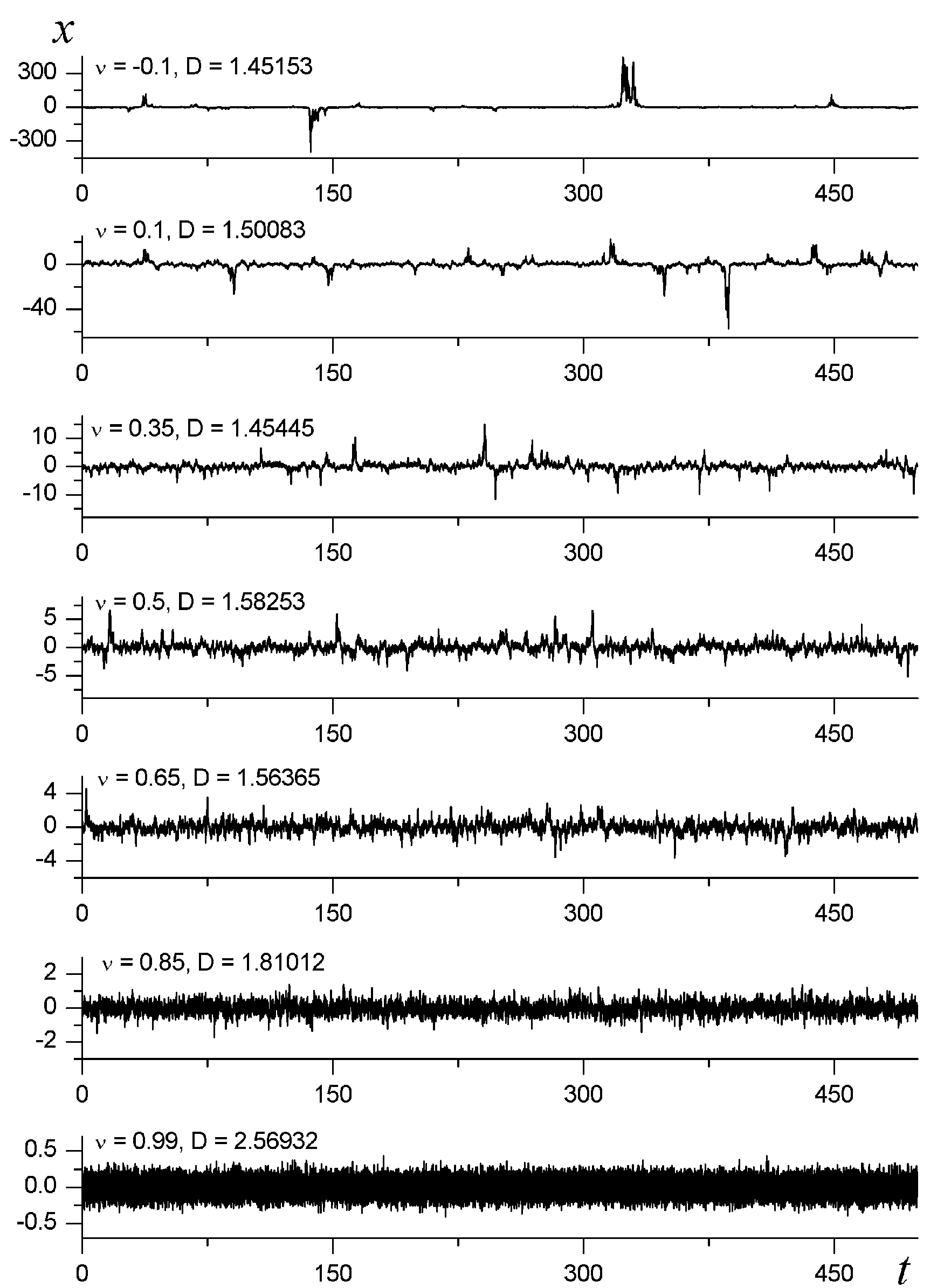} \caption{The form of modeled time series at ${\mathcal T}=500$,
$\tau=0.01$ and different $\nu$. The magnitudes of fractal dimension $D$ are
pointed out being determined in correspondence with $R/S$--method
\cite{Feder}.}
\end{figure}
As is seen from these series, at friction coefficient $-1<\gamma$, when the
parameter $\nu$ is negative, time series are relevant explicitly to the L\`evy
flights. With passage to positive parameters $\gamma$, $\nu$, their growth
arrives at gradual transformation of superdiffusion process into Brownian
diffusion that is related to the madnitudes $\gamma=\infty$, $\nu=1$. As $\nu$
increase, the relevant fractal dimension grows from the value $D\geq 1$ at
$\nu<0$ to $D\leq 2$ at $\nu\leq 1$.

In Figure 3, we show velocity time dependencies $v(t)$ obtained from origin
time series $x(t)$ according to definition (\ref{A}) where the simplest
non--clustering case $m=1$ is taken.
\begin{figure}[htb]
\centering
\includegraphics[width=130mm]
{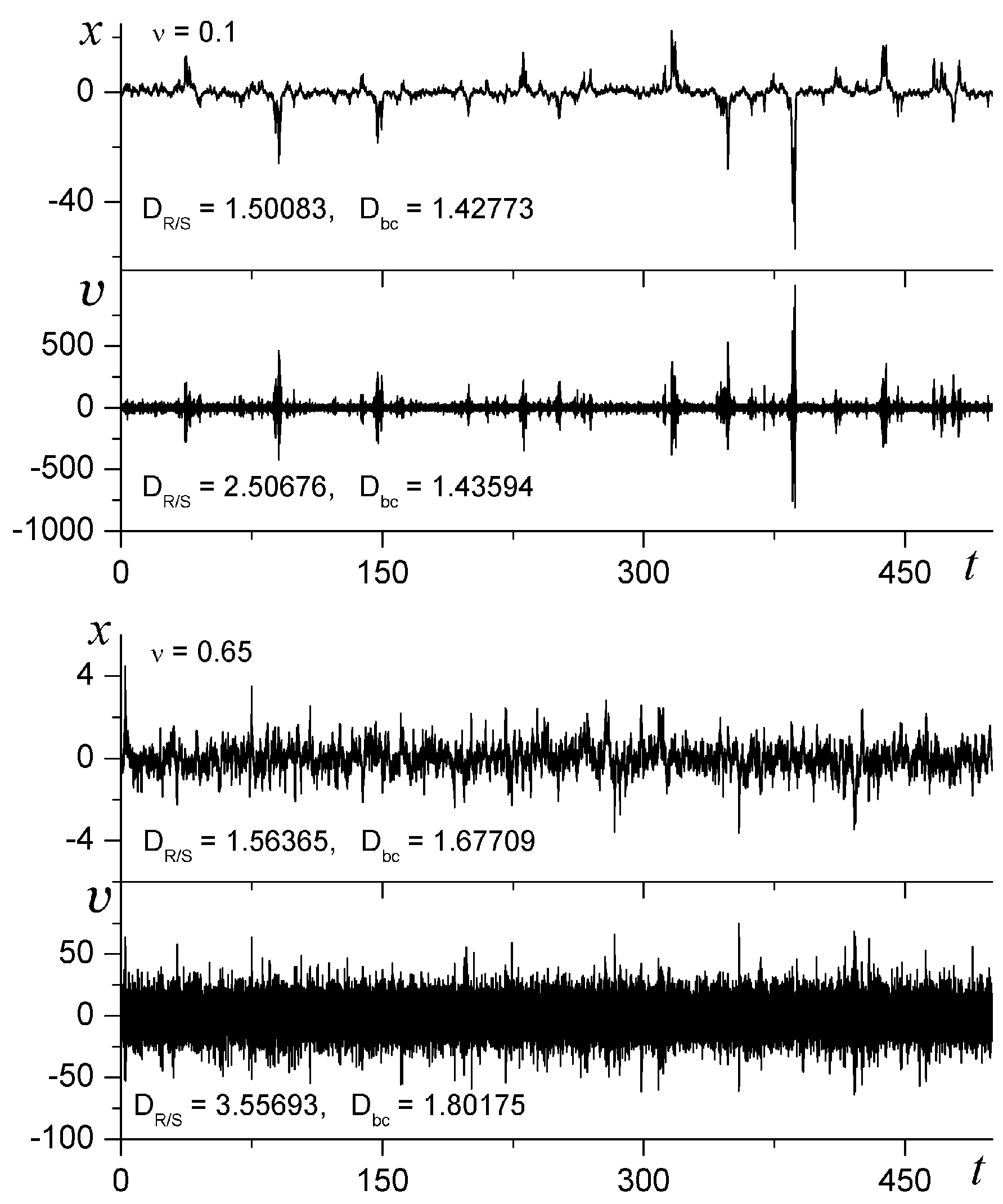} \caption{The form of velocity time dependencies $v(t)$ related to
time series $x(t)$. Because $R/S$--method gives overestimated values $D_{R/S}$
near upper boundary $D=2$ of fractal dimension, we apply also the values
$D_{bc}$ determined within box--counting method \cite{Feder}.}
\end{figure}
It is easily to see a velocity time dependence $v(t)$ has much more rugged form
in comparison with related time series $x(t)$. As a result, fractal dimension
for the velocity series $v(t)$ is always more than its value for the initial
series $x(t)$ (see data in Figure 3). Moreover, it turned out that $R/S$--method
and box--counting method (see \cite{Feder}) give quite different values
$D_{R/S}$ and $D_{bc}$ of fractal dimension. In our opinion, this difference
which grows with passage from time series $x(t)$ to velocity $v(t)$ is caused
by that related manifolds are more likely multifratals than fractals being
corresponded with single value of the fractal dimension.

\subsection{Statistical consideration of simulated time series}

To study statistical properties of above time series we are stated on the
ergodic hypothesis that supposes identity of averaging over both velocity time
series $v(t_i)$, $i=1,2,...,N_0$ and statistical ensemble $\{v_n\}$,
$n=1,2,...,N$, which describes velocity scattering in phase space. We determine
such an ensemble dividing maximum interval of the velocity variation domain
into $N\gg 1$ zones $n=1,2,...,N$, within of which velocities have mean value
$v_n$ and small variation $\delta v_n$. Then, the probability $p_n$ to hit the
interval $[v_n-\delta v_n/2,~v_n+\delta v_n/2]$ and the relevant probability
density function $\pi(v_n)$ are determined as follows:
\begin{equation}\label{pdf}
p_n\equiv\frac{\nu_n}{N_0}, \quad
\pi(v_n)\equiv N_0^{-1}\frac{\nu_n}{\delta v_n}
\end{equation}
where $\nu_n$ is a number of the time series specimens with mean value $v_n$.
According to Figure 4 the probability density function $\pi(v)$ has the usual
\begin{figure}[htb]
\centering
\includegraphics[width=80mm]
{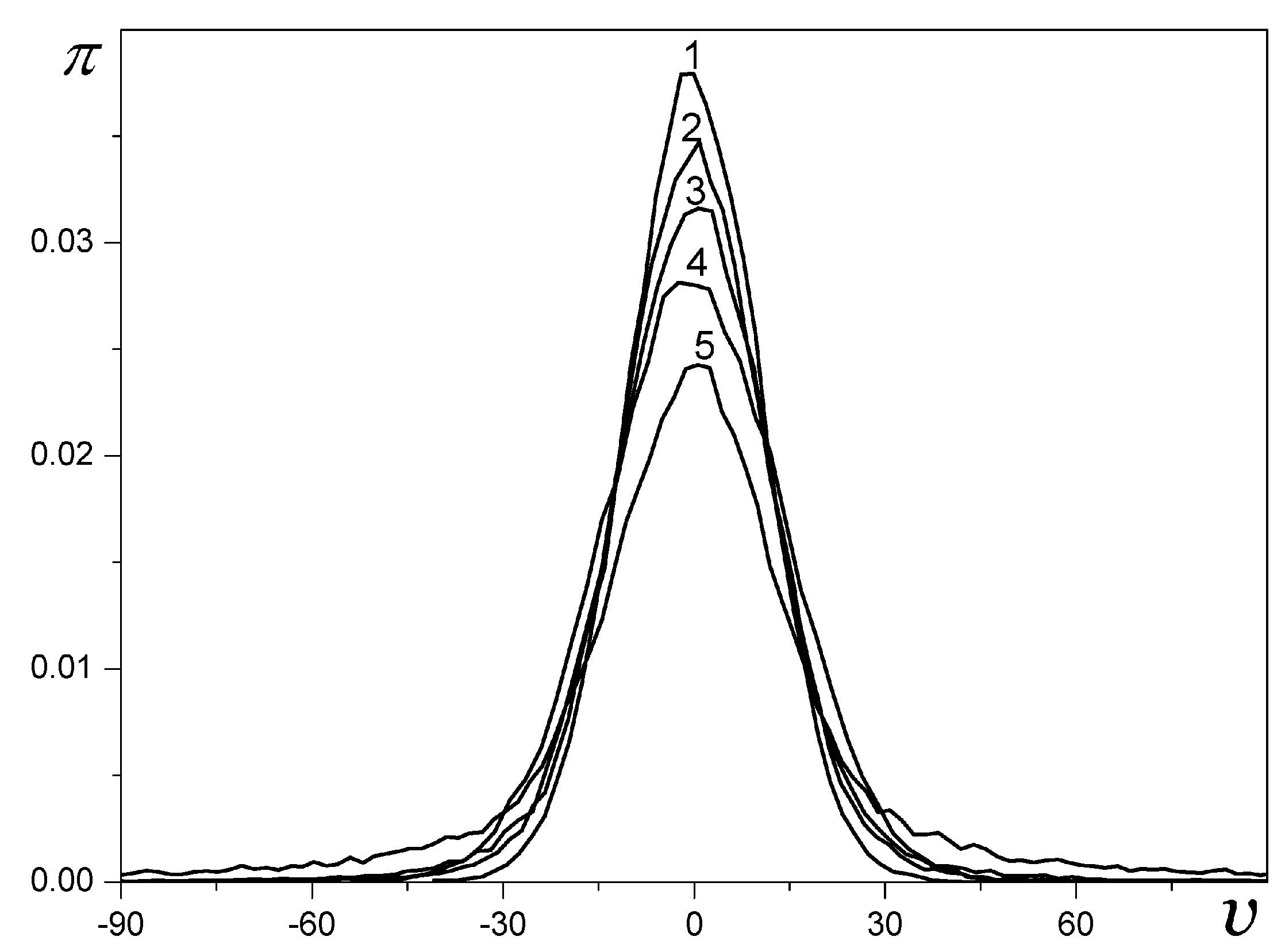} \caption{The probability density function of the velocity
distribution at different fractal dimensions (curves 1 --- 5 correspond to
parameters $\nu=0.85;~0.50;~0.35;~0.99;~-0.10$; respectively, fractal
dimensions are $D=1.81;~1.58;~1.45;~2.57;~1.45$).}
\end{figure}
bell--shaped form centered at the velocity $v=0$. By this, non-monotonic
variation of the $\pi(v)$ form in dependence of the fractal dimension $D$
attracts attention to oneself: in particular, the most broad velocity
scattering (curve 5) corresponds to negative value of the parameter $\nu$
to display fat tails.

Because distribution (\ref{pdf}) corresponds to grand canonical ensemble,
related thermodynamic functions are determined by the $n$--state number
\begin{equation}\label{num}
N_n\equiv Np_n=\frac{N}{N_0}\nu_n
\end{equation}
and energy (\ref{C}). Thus, the total energy of ideal gas\footnote{Do
not confuse with Hamiltonian (\ref{B}).} reads:
\begin{equation}\label{ene}
E\equiv\sum\limits_{n=1}^{N}\varepsilon_n
N_n=\frac{N}{N_0}\sum\limits_{n=1}^{N}\frac{v_n^2}{2}\nu_n.
\end{equation}
The plots of dependencies $E(N)$ of energy (\ref{ene}) accompanied with related
dependencies $H(N)$ of entropy (\ref{001}) are shown in Figures 5 for different
values of fractal dimensions $D$.
\begin{figure}[htb]
\centering
\includegraphics[width=130mm]
{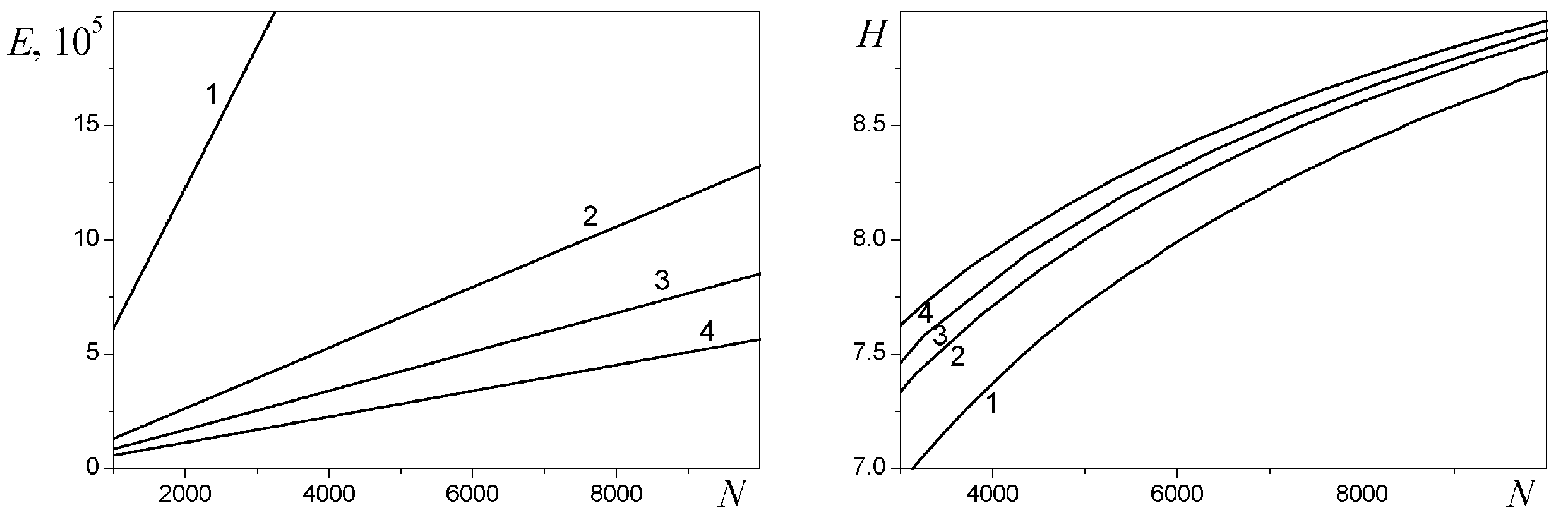} \caption{Dependencies of energy $E$ and entropy $H$ on the number
$N$ of effective particles (curves 1 --- 4 correspond to fractal dimensions
$D=1.50;~1.45;~1.58;~1.81$).}
\end{figure}
It is principally important, the energy $E$ is directly proportional to
the particle number $N$, while the entropy $H$ increases much more slowly
with $N$. As for the velocity distribution $\pi(v)$, non--monotonic
variations of both energy and entropy as functions of the fractal dimension
take place, but main tendency is
$E$ decrease and $H$ increase with $D$ growth.

\subsection{Calculations of effective temperatures of time series}

The peculiarity of the self--similar system under consideration is that it is
finite to be characterized with the following effective temperatures:
\begin{equation}\label{TT}
\Theta^{-1}\equiv \frac{\partial{H}}{\partial{E}}\Bigg|_D,\qquad T ^{-1}\equiv
N\frac{\partial{H}}{\partial{E}}\Bigg|_N.
\end{equation}
Being determined at constant value of the fractal dimension $D$, the first of
these is a function of the particle number $N$, whereas the second
one depends on magnitude $D$ to be determined at fixed value $N$. As show
simple calculations in Appendix B, above temperatures are connected by the
following relation:
\begin{equation}\label{ND}
T =\frac{\Theta}{N}\left[1-\frac{1}{(1-q)D^2}\left(\frac{H}{D-1}+
\frac{D-1}{D^2}N\ln N\right)^{-1}\right]^{-1}.
\end{equation}
Taking into account definitions (\ref{L1}), (\ref{Q}) and (\ref{1}) accompanied
with Eq.(\ref{Zz1}) where $T_s$ rescaled by factor $N$, one obtains
\begin{equation}\label{NDD}
T =\frac{\Theta}{N}\left\{1-\left[D\ln\left(\frac{NT
}{T_s}\right)+\frac{2}{d}\frac{\left(D-1\right)^2}{D}\ln N
\right]^{-1}\right\}^{-1}.
\end{equation}

Making use of the data presented in Figure 5 gives for the first temperature
(\ref{TT}) the dependence $\Theta(N)$ shown in the left panel of the Figure 6.
\begin{figure}[htb]
\centering
\includegraphics[width=130mm]
{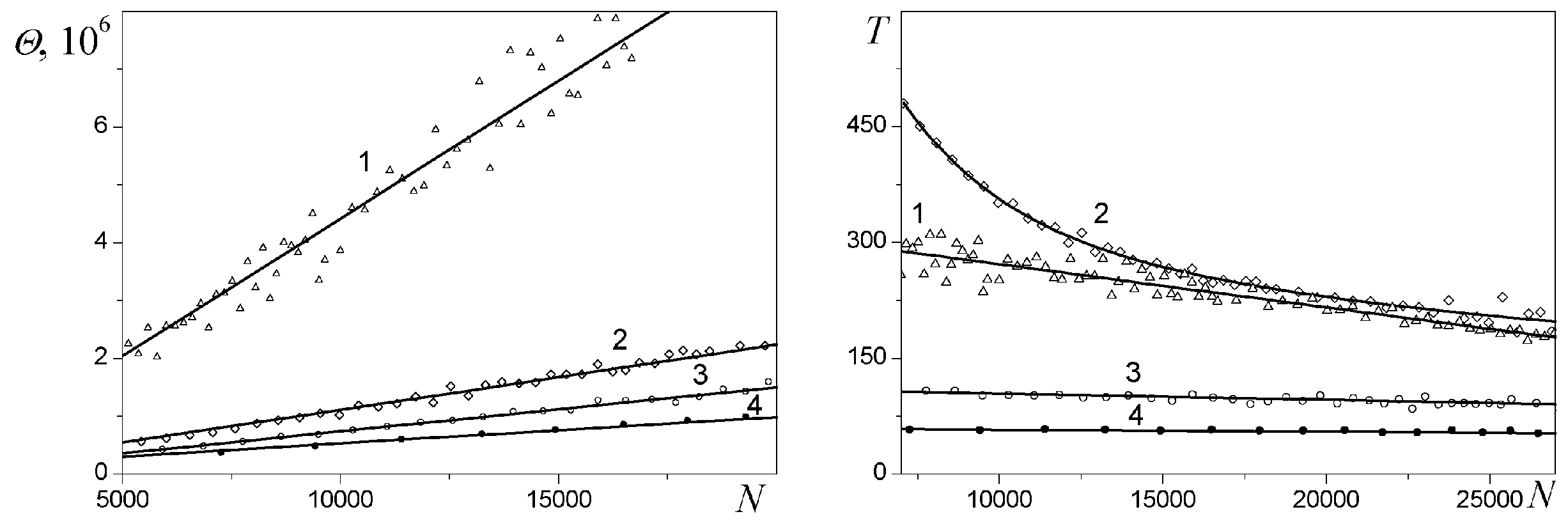} \caption{Dependencies of effective $\Theta$ and physical $T$
temperatures on the number $N$ of effective particles (curves 1 --- 4
correspond to fractal dimensions $D=1.50;~1.45;~1.58;~1.81$).}
\end{figure}
It is seen this temperature takes very large values which increase directly
proportionally with the $N$ number growth. Such a behavior displays an
intermediate character of the quantity $\Theta$ that plays rather a role of
effective energy being complementary to the energy (\ref{ene}). On the other
hand, usage of the relation (\ref{NDD}) arrives at the dependencies of the
temperature $T(N)$ shown in the right panel of the Figure 6. It is seen
with increase of the fractal dimension to values $D$ which are close to $D=2$
the dependencies $T(N)$ approach to constant magnitudes. These dimensions
relates to large values of the parameter $\nu\leq 1$ which is connected with
non--extensivity index $q$ according to equality $q=(2-\nu)^{-1}$. Thus, we
find the temperature $T$ takes values which are non--dependent on the
particle number $N$ in region $q\approx 1$ that is relevant to slightly
non--extensive limit where above analytical consideration is valid.
Due to such a behavior we can conclude the temperature $T$ has usual
physical sense.

To confirm this conclusion we examine the equipartition law (\ref{1}) rewritten
with usage of the averaged kinetic energy (\ref{C}):
\begin{eqnarray}
T=\langle\varepsilon\rangle\equiv\frac{\langle v^2\rangle}{2}.
\label{2341}
\end{eqnarray}
With this aim we compare the values $\langle\varepsilon\rangle$ and $T$ at
different magnitudes of the fractal dimension $D$ in Figure 7. It is seen the
\begin{figure}[htb]
\centering
\includegraphics[width=80mm]
{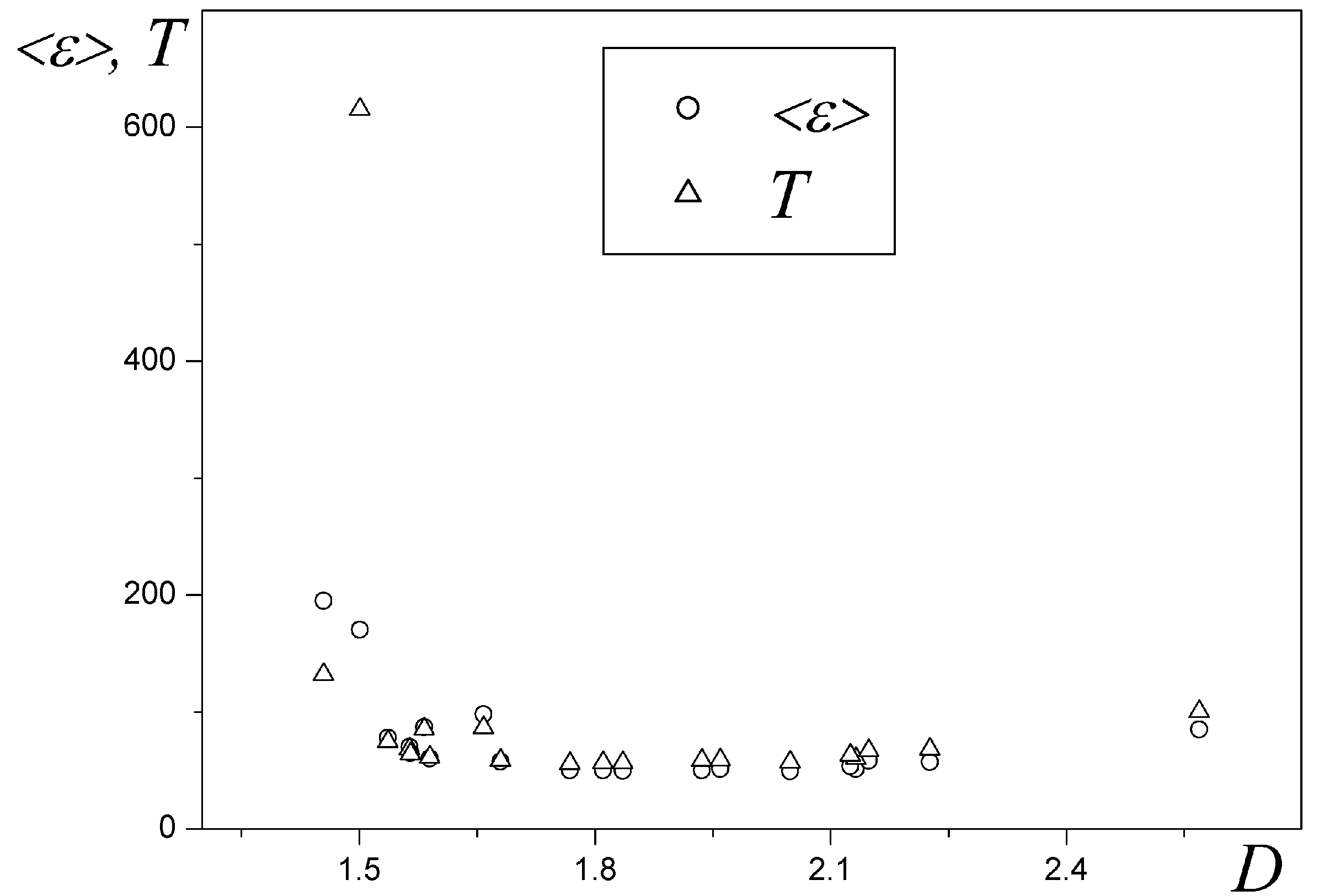} \caption{Dependencies of the averaged kinetic energy
$\langle\varepsilon\rangle$ and
the physical temperature $T$ on the fractal dimension $D$.}
\end{figure}
difference between above values is not exceed $10\%$ if one does not take into
account the points related to three smallest magnitudes of $D$ where our
analysis is not applicable.

\section{Discussion}

As shows above consideration, the quantity $T$ defined with the second equation
(\ref{TT}) plays the role of the physical temperature that determines the
energy (\ref{2341}) per one particle of slightly non--extensive ideal gas.
According to connection (\ref{28}) this temperature should grow monotonically
with increase of the fractal dimension, whereas Figure 7 shows non--monotonic
dependence. As is seen from Figure 2, such a behavior is caused by strong
falling down of the maximal variation $X$ with $D$ increase. To examine
exponential relation in Eq.(\ref{28}) we plot in Figure 8 the dependencies of
\begin{figure}[htb]
\centering
\includegraphics[width=130mm]
{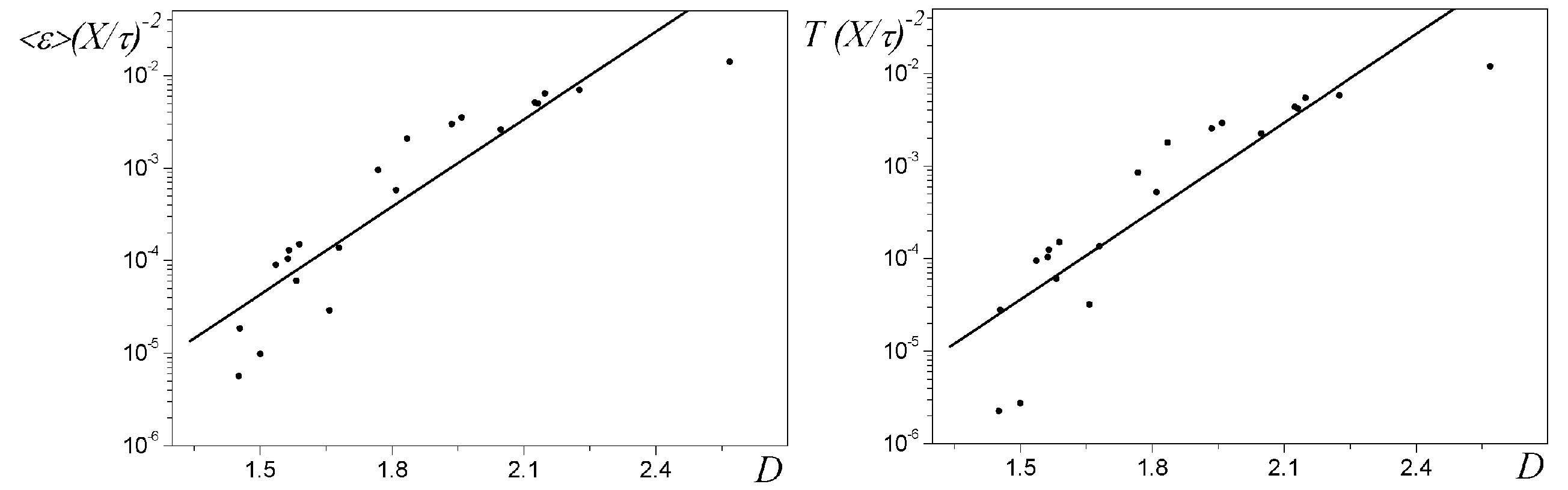} \caption{Graphical check--up of the dependence (\ref{28}): left
panel --- for mean value of kinetic energy $\langle\varepsilon\rangle$;
right panel --- for the physical temperature $T$.}
\end{figure}
the averaged special energy $\langle\varepsilon\rangle$ and the temperature $T$
on the fractal dimension $D$ using the factor $(X/\tau)^{2}$ as the scale of
these quantities. It is seen with such a scaling above dependencies take
approximately exponential form. Thus, main result of our analytical
consideration is conformed numerically.

It is worthwhile to pay attention to essential scattering of the numerical data
presented in Figures 6 --- 8. In our opinion, there are two main reasons of
above scattering. Firstly, this is caused by multifractal nature of
non--extensive time series generated according to recurrent procedure
(\ref{discret}). To take into account this reason it needs naturally to
consider a multifractional set whose fractal dimensions are distributed as a
function of the parameter $q=(2-\nu)^{-1}$ in accordance with the constraint
\begin{eqnarray}
D=\left(1-\frac{dN}{2}\cdot\frac{1-\nu}{2-\nu}\right)^{-1}. \label{Dnu}
\end{eqnarray}
that follows from Eqs.(\ref{L1}), (\ref{Q}). The second reason of the pointed
out scattering is clustering of time series that is displayed essentially at
small values of the parameter $\nu$ (see Figure 2). This reason is taken into
consideration by means of introducing effective interaction as it is done in
Subsection 2.3.

\section*{Appendix A: Main statements of the non--extensive
statistics}

Our approach is stated on using entropy definitions
\begin{eqnarray}
H=a\ln[\exp_q(H_q)],\quad H_q=\ln_q[\exp(H/a)];\qquad
a\equiv{1\over 2}(1-q)dN
\label{Aa}
\end{eqnarray}
that are alternations of the usual functions logarithm $\ln(x)$ and
exponential $\exp(x)$ with corresponding Tsallis generalizations
\cite{2}
\begin{eqnarray}
\ln_q(x)\equiv{x^{1-q} - 1\over 1-q},\quad
\exp_q(x)\equiv[1+(1-q)x]^{1\over 1-q}.
\label{00}
\end{eqnarray}
In explicit form, relations (\ref{Aa}) appears as
\begin{eqnarray}
H\equiv a\ln Z,\quad
H_q\equiv{\left< 1\right>_q - 1\over 1-q}.
\label{H1}
\end{eqnarray}
Making use of the first equality (\ref{Aa}) and the second formulae (\ref{00})
and (\ref{H1}) shows that physically defined entropy $H$ is extensive value
being reduced to the Renyi definition (\ref{000}):
\begin{eqnarray}
H=aK_q=\frac{Nd}{2}\ln\left<1\right>_q.
\label{001}
\end{eqnarray}
Comparison of this result with the first of equalities (\ref{H1})
arrives at the relation
\begin{eqnarray}
\left<1\right>_q=Z^{1-q}
\label{G}
\end{eqnarray}
that ensures the normalization condition \cite{2}.

\section*{Appendix B: Relationship between effective temperatures corresponding
to constant values of fractal dimension $D$ and particle number $N$}

Within determinant representation, the first of definitions (\ref{TT}) reads as
follows \cite{LL}
\begin{equation}\label{det}
\begin{split}
\Theta^{-1}=\frac{\partial(H,D)}{\partial(E,D)}=
\frac{\partial(H,D)/\partial(H,N)}{\partial(E,D)/\partial(H,N)} =\frac{\partial
D}{\partial N}\Bigg|_H\cdot\left[\det\Bigg(
\begin{array}{cc}
\frac{\partial E}{\partial H}\big|_N & \frac{\partial E}{\partial N}\big|_H \\
\frac{\partial D}{\partial H}\big|_N & \frac{\partial D}{\partial N}\big|_H
\end{array}
\Bigg)\right]^{-1}\\=\left[\frac{\partial E}{\partial H}\Bigg|_N-\frac{\partial
E}{\partial N}\Bigg|_H\cdot\frac{\partial D}{\partial
H}\Bigg|_N\cdot\left(\frac{\partial D}{\partial
N}\Bigg|_H\right)^{-1}\right]^{-1}.
\end{split}
\end{equation}
Making use of Eqs.(\ref{L1}), (\ref{Q}), (\ref{1}) gives
\begin{equation}\label{Det}
\frac{\partial D}{\partial N}\Bigg|_H=\frac{d}{2}(1-q)D^2,\qquad \frac{\partial
E}{\partial N}\Bigg|_H=\frac{d}{2}~T .
\end{equation}
Respectively, rewriting relation (\ref{P}) in the form
\begin{equation}
S=N\left[-\ln (gN)+\left(D\ln g +\frac{1}{D}\ln N\right)\right],\ \ g\equiv
G^{1/D}=(2\pi \mathrm{e} T_s)^{\frac{d}{2}}\left(\frac{x}{\tau}\right)^{-d},
\end{equation}
one finds
\begin{equation}\label{end}
\begin{split}
\frac{\partial H}{\partial D}\Bigg|_N=N\ln g-\frac{N}{D^2}\ln N
=N\left[\frac{1}{D}\ln\left(\frac{G}{N}\right)+\frac{1}{D}\ln
N-\frac{1}{D^2}\ln N\right]\\
=N\left[-\frac{H}{N}+\ln\left(\frac{G}{N}\right)+\frac{D-1}{D^2}\ln N\right]
=\frac{H}{D-1}+\frac{D-1}{D^2}N\ln N.
\end{split}
\end{equation}
Inserting Eqs.(\ref{TT}), (\ref{Det}), (\ref{end}) in last relation (\ref{det}),
one obtains the result (\ref{ND}).

\end{document}